\documentclass[aps,prl,twocolumn,nofootinbib]{revtex4}
\usepackage{graphicx}
\usepackage{bm}
\begin{document}
\newcommand{\postscript}[2]{\setlength{\epsfxsize}{#2\hsize}
   \centerline{\epsfbox{#1}}}
\newcommand{\mweak}{M_{\text{Weak}}}
\newcommand{\maux}{M_{\text{aux}}}
\newcommand{\msusy}{M_{\text{SUSY}}}
\newcommand{\mgut}{M_{\text{GUT}}}
\newcommand{\mplanck}{M_{\text{Pl}}}
\newcommand{\mstar}{M_{\ast}}
\newcommand{\md}{M_D}
\newcommand{\mbh}{M_{\text{BH}}}
\newcommand{\mbhmin}{M_{\text{BH}}^{\text{min}}}
\newcommand{\sbh}{S_{\text{BH}}}
\newcommand{\rbh}{R_{\text{BH}}}
\newcommand{\xmin}{x_{\text{min}}}
\newcommand{\ifb}{\text{fb}^{-1}}
\newcommand{\ev}{\text{eV}}
\newcommand{\mev}{\text{MeV}}
\newcommand{\gev}{\text{GeV}}
\newcommand{\tev}{\text{TeV}}
\newcommand{\pb}{\text{pb}}
\newcommand{\cm}{\text{cm}}
\newcommand{\km}{\text{km}}
\newcommand{\g}{\text{g}}
\newcommand{\s}{\text{s}}
\newcommand{\yr}{\text{yr}}
\newcommand{\sr}{\text{sr}}
\newcommand{\cmwe}{\text{cmwe}}
\newcommand{\kmwe}{\text{kmwe}}
\newcommand{\xmax}{X_{\text{max}}}
\newcommand{\etal}{{\em et al.}}
\newcommand{\eg}{{\em e.g.}}
\newcommand{\ibid}{{\em ibid.}}
\newcommand{\eqref}[1]{Eq.~(\ref{#1})}
\newcommand{\tbh}{\tau_{_{\rm BH}}}
\newcommand{\bold}[1]{{\text{\normalsize\boldmath $#1$}}}
\newcommand{\rem}[1]{{\bf #1}}
\newcommand{\lir}{\Lambda}
\newcommand{\be}{\begin{equation}}
\newcommand{\ee}{\end{equation}}
\title{Black Hole Chromosphere at the LHC}
\author{Luis Anchordoqui
and Haim Goldberg}
\affiliation{Department of Physics, Northeastern University, Boston, MA 02115}


\begin{abstract}

If the scale of quantum gravity is near a TeV, black holes will
be copiously produced at the LHC. In this work we study the main
properties of the light descendants of these black holes. We show
that the emitted partons are closely spaced outside the horizon,
and hence they do not fragment into hadrons in vacuum but more
likely into a kind of quark-gluon plasma. Consequently, the thermal
emission occurs far from the horizon, at a temperature
characteristic of the QCD scale. We analyze the energy spectrum
of the  particles emerging from the "chromosphere", and find that
the hard hadronic jets are almost entirely  suppressed. They are
replaced by an isotropic distribution of soft photons and
hadrons, with hundreds of particles in the GeV range. This
provides a new distinctive signature for black hole events at LHC.

\end{abstract}


\maketitle

{\em General idea:} In the mid-70's Hawking~\cite{Hawking:rv}
pointed out that a black
hole (BH) emits thermal radiation as if it were a black body,
with a temperature $T_{\rm BH}$ inversely proportional to its
mass $\mbh$. More recently,  Heckler~\cite{Heckler:1997jv} noted
that for $T_{\rm BH} > \Lambda_{\rm QCD}$, the emitted partons do
not immediatly fragment into hadrons, but propagate away from the
BH through a dense quark-gluon plasma losing energy via QCD
bremsstrahlung and pair production, and forming a nearly thermal
chromosphere. Assuming that TeV scale gravity is realized in
nature~\cite{Arkani-Hamed:1998rs}, in this work we first show
that such chromosphere would be shielding the BHs to be produced
at the Large Hadron Collider
(LHC)~\cite{Giddings:2001bu,Dimopoulos:2001hw},  and then we
estimate the emergent thermal photon spectrum from $\pi^0$ decay.
The ensuing discussion will be framed in the usual context of
TeV-scale gravity scenarios, in which the standard model (SM)
fields are confined to the brane and only gravity spills into the
$n$ internal dimensions of the universe. We also rely on the
probe brane approximation, {\it viz.}, the only effect of the
brane field is to bind the BH to the brane. This will be an
adequate approximation provided $\mbh$ is well above the brane
tension, which is presumably of the order of but smaller than the
fundamental Planck scale, $\md$. Moreover, we assume the BH can be
treated as a flat $4{+}n$ dimensional object. This assumption is
valid for both extra dimensions that are larger than the
Schwarzschild radius~\cite{Myers:un},
\begin{equation}
\label{schwarz}
r_s(\mbh) =
\frac{1}{\md}
\left[ \frac{\mbh}{\md} \,\, \frac{2^n \pi^{(n-3)/2}\Gamma({n+3\over 2})}{n+2}
\right]^{1/(1+n)}\,,
\end{equation}
and for warped extra dimensions where the
horizon radius is small compared to the curvature scale of the
geometry associated with the warped subspace~\cite{Giddings:2000ay}.

{\em Hawking evaporation:} According to the Thorne's hoop conjecture~\cite{Thorne:ji},
an event horizon forms when and only when a mass $\mbh$ is compacted into a region whose
circunference in every direction is less than $2 \pi r_s$. The strong gravitational
fields around the BH induce spontaneous creation of pairs near the event
horizon~\cite{Hawking:rv}. While the particle with positive energy can escape to infinity,
the one with negative energy has to tunnel through the horizon into
the BH where there are particle states with negative energy with respect to
infinity~\cite{Hartle:tp}. As the BHs radiate, they lose mass and so will eventually evaporate
completely and disappear. The evaporation is generally regarded as thermal in
character, with a temperature
\begin{equation}
T_{\rm BH} = \frac{n +1}{4\,\pi\,r_s}\,,
\end{equation}
and an entropy $S = 4\, \pi\,\mbh\,r_s/(n+2).$ However, the BH
produces an effective potential barrier in the neighborhood of
the horizon that backscatters part of the outgoing radiation,
modifing the  blackbody spectrum. The BH absorption cross
section, $\sigma_s$ (a.k.a. the greybody factor), depends upon
the spin of the emitted particles $s$, their energy $Q$, and the
mass of the BH $\mbh$~\cite{Page:df,Kanti:2002nr}. At high
frequencies ($ Q r_s \gg 1$) the greybody factor for each kind of
particle must approach the geometrical optics limit. For
illustrative simplicity, we adopt throughout this work the
geometric optics approximation, and following~\cite{Han:2002yy},
we conveniently write the greybody factor as a dimensionless
constant, $\Gamma_s = \sigma_s/A_4$, normalized to the BH surface
area
\begin{equation}
A_4 = 4\,\pi\,\left( \frac{n+3}{2} \right)^{2/(n+1)}\, \frac{n+3}{n+1} \, r_s^2
\end{equation}
seen by the SM fields ($\Gamma_{s=0} = 1$, $\Gamma_{s=1/2}
\approx 2/3$, and $\Gamma_{s=1} \approx 1/4$~\cite{Han:2002yy}).
The prevalent energies of the decay quanta are $\sim T_{\rm BH}
\sim 1/r_s$, resulting in $s$-wave dominance of the final state.
This implies that the BH decays with equal probability to a
particle on the brane and in the extra
dimensions~\cite{Emparan:2000rs}. Therefore, the evaporation
process is dominated by the large number of SM brane modes. Since
the $s$-wave greybody factors for fermions~\cite{Kanti:2002nr} differ from the geometric ones
by only about 10-20\%, we continue to use the latter.

A $4{+}n$ BH then emits
particles with initial total energy between $(Q, Q+dQ)$ at a rate
\begin{equation}
\frac{d\dot{N}_i}{dQ} = \frac{\sigma_s}{8 \,\pi^2}\,Q^2 \left[
\exp \left( \frac{Q}{T_{\rm BH}} \right) - (-1)^{2s} \right]^{-1}
\label{rate}
\end{equation}
per degree of particle freedom $i$. Integration of Eq.~(\ref{rate}) leads to
\begin{equation}
\dot{N_i} = f \frac{\Gamma_s}{32\,\pi^3} \,
\frac{(n+3)^{(n+3)/(n+1)}\,(n+1)}{2^{2/(n+1)}} \,\Gamma(3) \, \zeta(3) \,T_{\rm BH}\,,
\end{equation}
where $\Gamma(x)$ ($\zeta(x)$) is the Gamma (Riemann zeta) function and
$f=1$ ($f=3/4$) for bosons (fermions). Therefore, the BH emission rate is found to be
\begin{equation}
\dot{N}_i \approx 3.7 \times 10^{21}\, \frac{(n+3)^{(n+3)/(n+1)}}{2^{2/(n+1)} \,(n+1)^{-1}}\,
\left(\frac{T_{\rm BH}}{{\rm GeV}}\right)\,\, {\rm s}^{-1} \,\,,
\end{equation}
\begin{equation}
\dot{N}_i \approx 1.8 \times 10^{21}\, \frac{(n+3)^{(n+3)/(n+1)}}{2^{2/(n+1)} \,(n+1)^{-1}}\,
\left(\frac{T_{\rm BH}}{{\rm GeV}}\right)\,\, {\rm s}^{-1} \,\,,
\end{equation}
\begin{equation}
\dot{N}_i \approx 9.2 \times 10^{20}\,\frac{(n+3)^{(n+3)/(n+1)}}{2^{2/(n+1)}
\,(n+1)^{-1}}\,
\left(\frac{T_{\rm BH}}{{\rm GeV}}\right)\,\, {\rm s}^{-1} \,\,,
\end{equation}
for particles with $s = 0,\, 1/2, \,1,$ respectively. Note that the BHs to be
produced at the LHC are perfectly well defined resonances with mean lifetimes,
$\tau_{_{\rm BH}} \sim M^{-1}_D \,(\mbh/M_D)^{(n+3)/(n+1)} \approx 10^{-27}$~s, and $T_{\rm BH} \approx 200$~GeV.

Since thermal fluctuations due to particle emission are small
when the entropy $S \gg 1$~\cite{Preskill:1991tb} and statistical
fluctuations in the microcanonical ensemble are small for
$\sqrt{S} \gg 1$~\cite{Giddings:2001bu}, we expect the above
formulae to be an adequate approximation for $\mbh \gg \md$. This
condition inevitably breaks down during the last stages of the
decay process. However, it is noteworthy that for BHs with
initial masses well above $M_D$ most of the evaporation takes
place within the semi-classical regime. In what follows we
require $\mbh/\md \agt 5$, corresponding to $S \agt 25$. To
obtain some quantitative estimates on the particle spectra, we
hereafter set $n = 6$ and $\md = 1.3$~TeV, a value consistent with
current limits on the size of
extra-dimensions~\cite{Abbott:2000zb}.

{\em Thermalization:} The evaporation of the BH creates a
radiation shell of radius $r_s$ and thickness $\tbh,$ which
propagates outward at the speed of light. Using the preceding
equations we find that, already for $\mbh/\md = 5$, there are
about 10 quarks and antiquarks in the shell. Before proceeding
further, we examine under what conditions these particles can
undergo sufficient dissipative interactions (bremsstrahlung and
pair production) to initiate thermalization.

\begin{figure}
\includegraphics[width=0.3\textwidth,height=3.5cm]{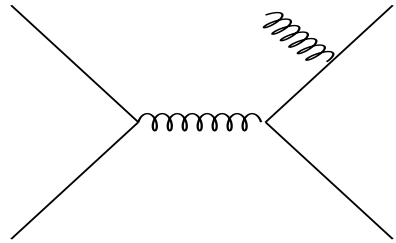}
\caption{Feynman diagram for the dominant contribution to gluon bremsstrahlung
in a $qq$ or $q \bar q$ collision.}
\label{brem}
\end{figure}

At any time $t$ after the end of evaporation, the shell has
radius $r=ct$ and thickness $\tbh.$ A quark or antiquark  in the
shell will interact at a rate given by
\begin{equation}
\Gamma = n\,\,c\,\,\sigma^{^{\rm QCD}}_{_{\rm brem}}\ \ ,
\label{rate+}
\end{equation}
where $n$ is the $q\bar q$ effective density at time $t$, and
$\sigma^{^{\rm QCD}}_{_{\rm brem}}$ is the cross section for gluon
bremsstrahlung in a $qq$ or $q\bar q$ collision. The dominant
diagram of this process at low momentum transfer is shown in
Fig.~\ref{brem}. The energy-averaged total cross section is given
by~\cite{Heckler:1997jv}
\begin{equation}
\sigma^{^{\rm QCD}}_{_{\rm brem}} \approx
\frac{8\,\alpha_s^3}{\Lambda^2}\
\ln\left(\frac{2\,Q}{\Lambda}\right)\ \ , \label{sigma}
\end{equation}
where $\alpha_s$ is the QCD coupling constant, and $\lir$ is an
infrared cutoff related to the off-shell momentum of the
exchanged gluon. Note that the exchanged (virtual) gluon can
travel at most a distance $\sim r$ during the interaction, so that the
uncertainty principle provides a lower bound $\sim  r^{-1}$ on the
momentum transfer. Thus, at any time $t=r/c,$  the cross section
has a maximum value corresponding to  $\lir\approx r^{-1}.$ Of
course, $\lir$ is absolutely bounded from below by $\Lambda_{\rm
QCD}.$

In light of these considerations, we take  the effective density
to be fully volumetric
\be n = \frac{N_{q \bar q}}{{4 \over
3}\,\pi\,r^3}\
\ee
so that a quark or antiquark interacts at a  rate
\begin{equation}
\Gamma =  \frac{6\, N_{q \bar
q} \, \alpha_s^3}{\pi}\, \ln\left[\frac{2Q}{\Lambda}\right]
\,\left( \frac{1}{\Lambda r} \right)^2 \,
\frac{1}{r}\,.\label{nint2}
\end{equation}
For $r\alt \Lambda_{\rm QCD}^{-1},$ we can choose $\Lambda
\approx r^{-1}.$ In order to remain perturbative, we find the
cumulative number of interactions/quark for $\Lambda\sim r^{-1}$
down to some value $\gg \Lambda_{\rm QCD}.$ Integration of
(\ref{nint2}) with $N_{q \bar q}\approx 10$ leads to \be {\cal
N}_{\rm int} \approx 0.15\
\left(\frac{\alpha_s(\Lambda)}{0.2}\right)^3
\ln\left[\frac{2Q}{\Lambda}\right]
\ln\left[\frac{1}{\Lambda\tbh}\right]\ \ .\label{realnint} \ee The
fiducial value of $\alpha_s$ has been chosen appropriate to a
momentum transfer scale of 9 GeV~\cite{Hagiwara:pw}, corresponding
to upsilon decay. Even for such a large value of
$\Lambda$
and $Q\simeq 2 \,T_{\rm BH} \simeq 400$ GeV,
Eq.~(\ref{realnint}) yields $\approx$ 3 interactions per quark;
for $\Lambda = 1.78 $ GeV, corresponding to $\tau$-decay~\cite{Hagiwara:pw},
$\alpha_s \simeq 0.35,$ and
Eq.~(\ref{realnint}) yields $\approx$ 30 interactions per quark.
The creation of a gluon component in the plasma will lead to a
significant increase in the bremsstrahlung cascade, because the
gluon-gluon cross section is considerably larger than the one of
quark-(anti)quark. We take this result as supportive of
thermalization, and we proceed on that assumption.

Before interactions have led to a rapid increase in particle numbers, the
average separation at distance $r$ of partons emitted by the BH
is defined as
\begin{equation}
\langle d\rangle_{\rm parton}  \equiv
 \left(\sum_{i = q, \bar q, g} \frac{c_i\,\,N_i}{{4 \over 3}\,\pi\,r^3}
\right)^{-1/3} \, ,
\label{d}
\end{equation}
where $c_i$ is the number of internal
degrees of freedom of particle species $i$, {i.e.}, 72  (16) for quarks
and antiquarks (gluons). Evaluating this expression
for our parameters, we find that $\langle d\rangle_{\rm parton}
\sim 7 \times 10^{-2} {\rm GeV}^{-1} \ll \Lambda_{\rm QCD}^{-1}.$ As interactions cascade,
the volume occupied by the partons expands as well. The temperature of the chromosphere
decreases as the
radius increases, and eventually   reaches
\begin{equation}
\langle d \rangle_{\rm parton} \sim \Lambda_{\rm QCD}^{-1} \,,
\end{equation}
at which point quarks and gluons hadronize.

{\em Soft thermal spectrum:} The density of the
outward-going plasma changes by a large factor within a mean free
path, and so the particles never have enough time to fully thermalize.
In order to describe the outer edge of the chromosphere we adopt here the
heuristic treatment given by Heckler~\cite{Heckler:1997jv}. The quark and
gluon spectrum in the observer rest frame is obtained by boosting a thermal
spectrum at $T_{_{\rm ch}} = \Lambda_{\rm QCD}$ with the Lorentz factor,
$\gamma_{_{\rm ch}} \approx (T_{\rm BH}/\Lambda_{\rm QCD})^{1/2}$, of the
outer
surface of the thermal chromosphere, i.e.,
\begin{eqnarray}
\frac{d\dot{N}_i}{dQ} & = & c_i \frac{\gamma_{_{\rm ch}}^2\,
r_{_{\rm ch}}^2 \, Q^2}{2 \pi^2}
\int_0^1  \,d\Omega \, (1 - \beta\, \cos \theta) \, \cos \theta \nonumber \\
& & \left\{ \exp \left[ \frac{\gamma_{_{\rm ch}} Q (1-\beta \cos
\theta)}{T_{_{\rm ch}}} \right] - (-1)^{2s} \right\}^{-1},
\label{qg}
\end{eqnarray}
where the integration over the surface of the chromosphere of radius $r_{_{\rm ch }} = \gamma_{_{\rm ch}}/\Lambda_{\rm QCD}$  has been carried out. It is
noteworthy that detailed analyses~\cite{Cline:1998xk} based on the
transport equation yield results in very good agreement with this spectrum.

The observed photon spectrum of the chromosphere is a convolution of the
quark-gluon spectrum given in Eq.~(\ref{qg}) with the pion fragmentation
function and the Lorentz-boosted spectrum from $\pi^0$ decay; namely,
\begin{equation}
\frac{d \dot{N}}{dE_{\rm \gamma}} = \int_{E_0}^\infty dE_\pi \,
\frac{dg_{\pi \gamma} (E_\pi)}{dE_{\gamma}} \frac{d \dot{N}_\pi}{dE_\pi}\,,
\label{13}
\end{equation}
where $E_0 = E_\gamma + m_\pi^2/ 4 E_\gamma$. The number of photons with
energy $E_\gamma$ produced by a pion propagating with velocity $\beta$ and
decaying isotropically in its rest frame is
\begin{equation}
\frac{dg_{\pi \gamma} (E_\pi)}{dE_\gamma} = \frac{2}{\gamma m_\pi \beta} =
\frac{2}{(E_\pi^2 - m_\pi^2)^{1/2}} \,,
\end{equation}
with $\gamma = (1 - \beta^2)^{-1/2}$. Finally, the pion spectrum
can be expressed in the form~\cite{MacGibbon:zk}
\begin{equation}
\frac{d \dot{N}_\pi}{dE_\pi} = \sum_j \int_{E_\pi}^\infty
\frac{d\dot{N}_j (Q,T_{_{\rm ch}})}{dQ}\, \frac{dg_{j\pi} (Q, E_\pi)}{dE_\pi} \, dQ \ ,
\end{equation}
where the sum is over relevant species in the plasma.
The precise nature of the fragmentation process is unknown. As
in our previous analysis of BH decay~\cite{Anchordoqui:2001ei}, we adopt
the quark $\rightarrow$ hadron fragmentation function originally
suggested by Hill~\cite{Hill:1982iq}
\begin{equation}
\frac{dg_{j\pi}}{dE_\pi} \approx \frac{15}{16} \, z^{-3/2} \, (1-z)^2\,,
\end{equation}
that is consistent with the so-called ``leading-log QCD'' behavior and seems
to reproduce quite well the multiplicity growth as seen in collider
experiments ($z= E_\pi/Q$).

Using Eq.~(\ref{13}) one can easily estimate the mean lifetime of the
chromosphere
\begin{equation}
\tau_{\rm ch} \approx \frac{\mbh}{3}\, \frac{1}{E_{\gamma_{\rm max}} \,
\Delta E_\gamma}\,  \left.
\frac{d\dot{N}}{d E_\gamma}\right|_{\rm max}^{-1} \,,
\end{equation}
where $\Delta E_\gamma$ is the full width at half maximum (max) of
the $d \dot{N}/dE_{\rm \gamma}$ distribution.
The spectrum peaks
at an energy of $E_{\gamma_{\rm max}} \approx 100$~MeV, and thus
for our choice of parameters one obtains $\tau_{\rm ch}
\approx 2.2 \times 10^2 \,\,\tau_{_{\rm BH}}\,.$

\begin{figure}
\includegraphics[width=0.5\textwidth,height=8cm]{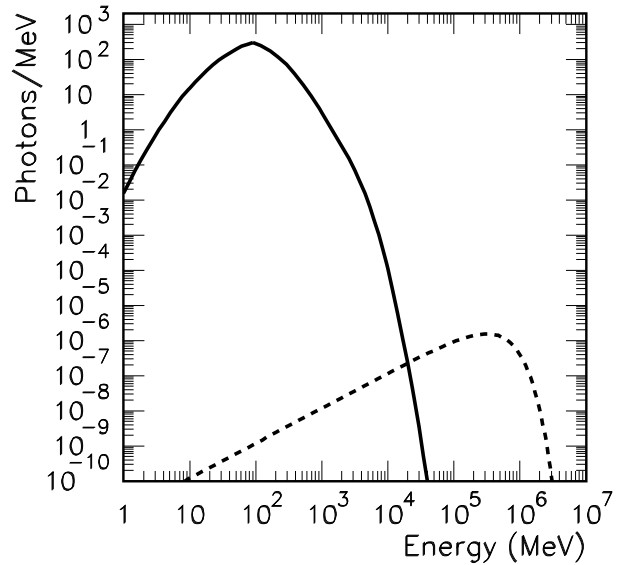}
\caption{Photon emission spectra of a 10-dimensional Schwarzschild BH,
assuming $\md \approx 1.3$~TeV and $\mbh/\md~=~5$. This implies,
$T_{\rm BH} \approx 200$~GeV and $\tau_{_{\rm BH}}
\approx 4 \times 10^{-27}$~s. The solid line denotes the
spectrum from quark fragmentation and subsequent $\pi^0$ decay, whereas the
dashed line stands for direct photon emission from the BH.}
\label{bh_lhc}
\end{figure}

In Fig.~\ref{bh_lhc} we show the emergent photon spectrum from the
chromosphere hadronization and subsequent $\pi^0$ decay, together with that
of direct photon emission from the BH. The photon emission rate from the
chromosphere is found to be~\cite{Heckler:1997jv}
\begin{equation}
\dot{N}_{\gamma_{\rm ch}} \approx 2 \times 10^{24} \,\left(\frac{T_{\rm BH}}{\rm GeV}\right)^2\,
\, {\rm s}^{-1}\,.
\end{equation}
Therefore, for a BH with $\tau_{_{\rm BH}} \approx 4 \times 10^{-27}$~s and
$T_{\rm BH} \approx 200$~GeV, one expects $\approx 7 \times 10^4$ photons
as end products of the quark-gluon plasma. The nominal Compact Muon Solenoid (CMS)
design anticipates the dynamic range for the electromagnetic calorimeter
readout will extend from the noise floor of 25~MeV up to 2 or 2.5~TeV,
whereas the hadron calorimeter readout will range from 20~MeV to
2~TeV~\cite{CMS}. Consequently, detection of chromosphere's radiation could
become observationally feasible. Moreover, a $\pi^{\pm}$ spectrum with
roughly the same shape, but shifted in energy by a factor of 2, would trigger
the hadron calorimeter.

In general, one expects a statistically significant signal
at the LHC from both  $pp$~\cite{Dimopoulos:2001hw} and PbPb
collisions~\cite{Chamblin:2002ad}. In particular, for the above parameters
and assuming a geometric cross section for the colliding partons,
about $10^{6}$ BH events with zero net transverse momentum $p_T$ would be
produced with an integrated luminosity
of 100~fb$^{-1}$~\cite{Dimopoulos:2001hw}. Proposed modifications,
such as BH produced with high $p_T$~\cite{Cheung:2001ue} and recoil effects
from graviton emission into the bulk~\cite{Frolov:2002as} do not
qualitatively change this estimate. Criticisms of the absorptive black disc
scattering amplitude,
which center on the exponential supression of transitions involving
a (few-particle) quantum state to a (many-particle) semiclassical
state~\cite{Voloshin:2001vs},
have been addressed in~\cite{Dimopoulos:2001qe}. However, it is
noteworthy that even if one includes this supression, the BH production
rate will still be quite large at the LHC for over almost all of interesting
parameter space~\cite{Rizzo:2002kb}. Specifically, for our fiducial values
about $10^{4}$ BHs (either at rest or travelling along the beam pipe) would
be produced with an integrated luminosity of 100~fb$^{-1}$.

Note that there is a mild dependence of $T_{\rm BH}$ with the
initial BH mass. Consequently, most of the BHs ($\mbh/M_D \agt
5$) to be produced at the LHC have temperatures $\approx
200$~GeV. The spectrum from quark fragmentation and subsequent
$\pi^0$ decay given in Fig.~\ref{bh_lhc} is thus quite general.

{\em Summary and final remarks:} The results of this work indicate
that the signals for BH production delineated by
Giddings--Thomas~\cite{Giddings:2001bu}
and Dimopoulos--Landsberg~\cite{Dimopoulos:2001hw} (large transverse momentum,
hard electromagnetic component) can be strikingly augmented with
the observation of the soft thermal isotropic hadronic and gamma
ray component, which includes significant photon and charged pion
counting rates in the vicinity of $\sim 1$~GeV. As can be seen
from Fig.~\ref{bh_lhc}, one expects about 100 photons (as well as
about 200 charged pions) per black hole event with energies in the
GeV range. {\em A hard hadronic component is largely absent.} This
is much more dramatic than the absence of hadronic jets with
$p_T>r_s^{-1}\sim T_{\rm BH}$~\cite{Banks:1999gd}. According to
calorimeter design guidelines for the CMS detector, the
observation of this component should be feasible.

TeV-scale BHs can also be produced in cosmic ray
collisions~\cite{Feng:2001ib}, and (of course) one expects the
quarks and gluons to be at first freely streaming away from the
horizon. However, contrary to collider experiments, the BH is
produced with large momentum in the lab system, and its decay
products are  swept forward with large, ${\cal O}(10^6),$ Lorentz factors. This
minimizes the effect of the chromosphere on the spectra of the
emergent particles analyzed in Ref.~\cite{Anchordoqui:2001ei}. In particular, a 100~MeV pion boosted 
by $10^6$ will still have a mean decay length to muons of $\sim 8000$~km, much
larger than 1 km, which is the estimated mean interaction length of
a pion of $10^5$~GeV~\cite{Anchordoqui:1998nq}. Thus, an extensive air shower
will still occur. The final detectability of the shower will differ
little from the situation when the decay particles have energies $\sim 100$~GeV. If 
anything, the chromosphere will accelerate the muon production, for reasons similar 
to the enhancement of the muon component in a shower initiated by a complex nucleus.

Our analysis impacts directly on the identification of the top
quark among the BH subproducts~\cite{Uehara:2002gv}. The decay
rate of the $t$--quark in the SM is approximately
\be
\Gamma_t \sim \frac{3}{16\pi}\, m_t \, m_W^2 \,(G_F/\sqrt{2}) \sim 0.5\,\,{\rm GeV} \,,
\ee
yielding a
lifetime $\tau_t \sim 2$~GeV$^{-1}$. Here, $m_t\ (m_W)$ denotes
the mass of the top ($W$) and $G_F$ is the Fermi coupling
constant. Now, since $r_{\rm ch}\approx T_{\rm
BH}^{1/2}/\Lambda_{\rm QCD}^{3/2}\approx 150$~GeV$^{-1}$ (and the
effects of time dilation are negligible), top quarks emitted as
part of  the Hawking radiation cannot escape the chromosphere
even in the absence of collisions. Diffusive effects will
increase the time spent in the chromosphere, and thus strengthen
this result. Therefore, top quark identification depends on the
leptonic activity resulting from their decay. We also note that both
the $W$ and $Z$ decay inside the chromosphere. On the other hand,
the lifetime for a Higgs boson decaying into $b\bar b$ is
\begin{eqnarray}
\tau_{H\rightarrow b\bar b} & = &
{1\over 3}\ (\cos\beta/\sin\alpha)^2\ \left(\frac{v/\sqrt{2}}{m_b}\right)^2\ \frac
{16\pi} {m_H} \nonumber \\
 &  \simeq & 140 \,\,\left(\frac{\cos\beta}{\sin\alpha}\right)^2 \,
 \left(\frac{150\ {\rm GeV}}{m_H}\right)\ {\rm GeV}^{-1}\ ,
\label{higgs}\end{eqnarray}
where $m_b\ (m_H$) is the mass of the $b$-quark (Higgs) and $v=246$ GeV.
In two-Higgs models (such as supersymmetry), $\tan\beta$ is the ratio of
the vevs and $\alpha$
is a mixing angle in the neutral Higgs sector. For the SM one-Higgs case, the factor
$\cos\beta/\sin\alpha$ is set equal to 1.
Thus, in the SM, the Higgs seems likely to escape the chromosphere and generate a hard
jetty component in small percentage of the BH decays~\cite{Landsberg:2001sj}.
In the two-Higgs model,
escape from the chromosphere will depend on the mixing angles.

\hfill

\begin{acknowledgments}
We are thankful to Steve
Giddings for insightful remarks on the manuscript.
We are also indebted to Tom Paul
for some valuable discussion on design considerations of the CMS
calorimeters. Helpful suggestions from the journal referee are gratefully
acknowledged. The work of LA and HG has been
partially supported by the US National Science Foundation (NSF), under
grants No.\ PHY--9972170 and No.\ PHY--0073034, respectively.
\end{acknowledgments}


\end{document}